%% file: monow.tex
\documentclass[aps,prl,nofootinbib,twocolumn,showpacs,superscriptaddress,letterpaper,amsmath,amssymb]{revtex4}
\usepackage{preprintcover}
\usepackage[hyperindex,breaklinks]{hyperref}
\usepackage{graphicx} 
\usepackage{dcolumn}  
\usepackage{bm}       
\usepackage{feynmf}
\usepackage{slashed}
\usepackage{multirow}
\def\gsim{\lower0.5ex\hbox{$\:\buildrel >\over\sim\:$}}
\def\lsim{\lower0.5ex\hbox{$\:\buildrel <\over\sim\:$}}

\newcommand{\be}{\begin{equation}}
\newcommand{\ee}{\end{equation}}
\newcommand{\bea}{\begin{eqnarray}}
\newcommand{\eea}{\end{eqnarray}}

\newcommand{\nbox}{{\,\lower0.9pt\vbox{\hrule \hbox{\vrule height 0.2 cm
\hskip 0.2 cm \vrule height 0.2 cm}\hrule}\,}}

\def\missET {E_{\textrm{T}}^{\textrm{miss}}}

\PreprintCoverPaperTitle{Search for dark matter in events with a hadronically decaying $W$ or $Z$ boson and missing transverse momentum in $pp$ collisions at $\sqrt{s}=8$~TeV with the ATLAS detector}
\PreprintIdNumber{CERN-PH-EP-2013-158}
\PreprintCoverAbstract{
  A search is presented for dark matter pair production in association
  with a $W$ or $Z$ boson in $pp$ collisions representing 20.3
  fb$^{-1}$ of integrated luminosity at $\sqrt{s}=8$~TeV using data
  recorded with the ATLAS detector at the Large Hadron
  Collider. Events with a hadronic jet with the jet mass consistent with a $W$ or $Z$
  boson, and with large missing transverse momentum are analyzed.
  The data are consistent with the standard model expectations. 
  Limits are set on the mass scale in effective field theories that describe the interaction of dark matter and standard model particles, and on the cross section of Higgs production and decay to invisible particles.
In addition, cross section limits on the anomalous production of $W$ or $Z$ bosons with large missing transverse momentum are set in two fiducial regions.
}
\PreprintJournalName{Physical Review Letters}

\begin{document}

\thispagestyle{empty}
\vspace*{-3.5cm}

\vspace{0.5in}

\title{Search for Dark Matter in Events with a Hadronically Decaying $W$ or $Z$ Boson and Missing Transverse Momentum in $pp$ Collisions at $\sqrt{s}=8$~TeV with the ATLAS Detector}

\begin{center}
\begin{abstract}
  A search is presented for dark matter pair production in association
  with a $W$ or $Z$ boson in $pp$ collisions representing 20.3
  fb$^{-1}$ of integrated luminosity at $\sqrt{s}=8$~TeV using data
  recorded with the ATLAS detector at the Large Hadron
  Collider. Events with a hadronic jet with the jet mass consistent with a $W$ or $Z$
  boson, and with large missing transverse momentum are analyzed.
The data are consistent with the standard model expectations. 
Limits are set on the mass scale in effective field theories that describe the interaction of dark matter and standard model particles, and on the cross section of Higgs production and decay to invisible particles.
In addition, cross section limits on the anomalous production of $W$ or $Z$ bosons with large missing transverse momentum are set in two fiducial regions.
\end{abstract}
\end{center}

\author{ATLAS Collaboration}

\maketitle

Although the presence of dark matter in the Universe is
well established, little is known of its particle nature or its
nongravitational interactions.  A suite of experiments is searching
for a weakly interacting massive particle (WIMP), denoted by
$\chi$, and for interactions between $\chi$ and standard model (SM)
particles~\cite{PhysRevD.86.010001}.

\begin{figure}[thbp]
\center
\includegraphics[width=0.4\linewidth]{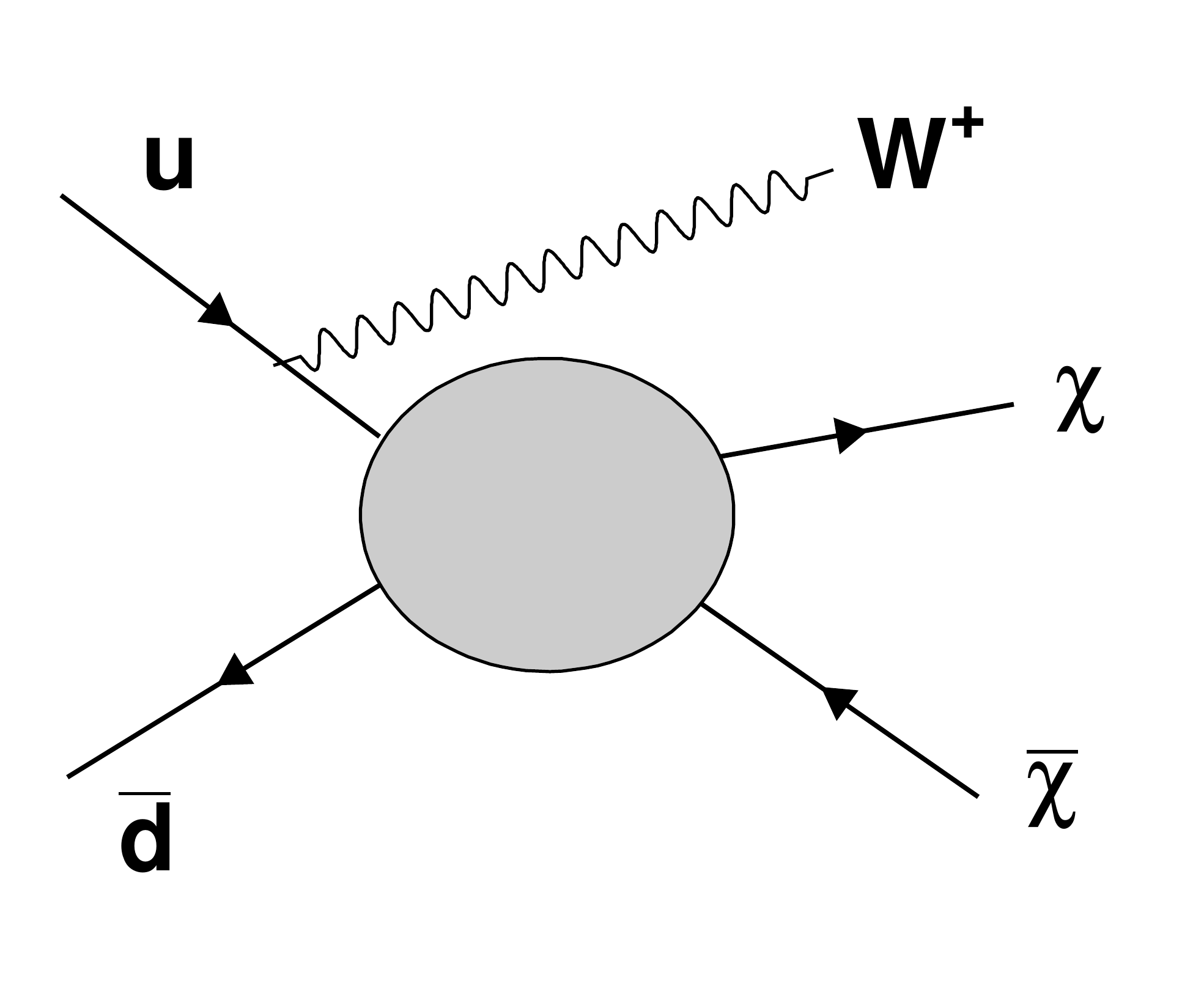}
\includegraphics[width=0.4\linewidth]{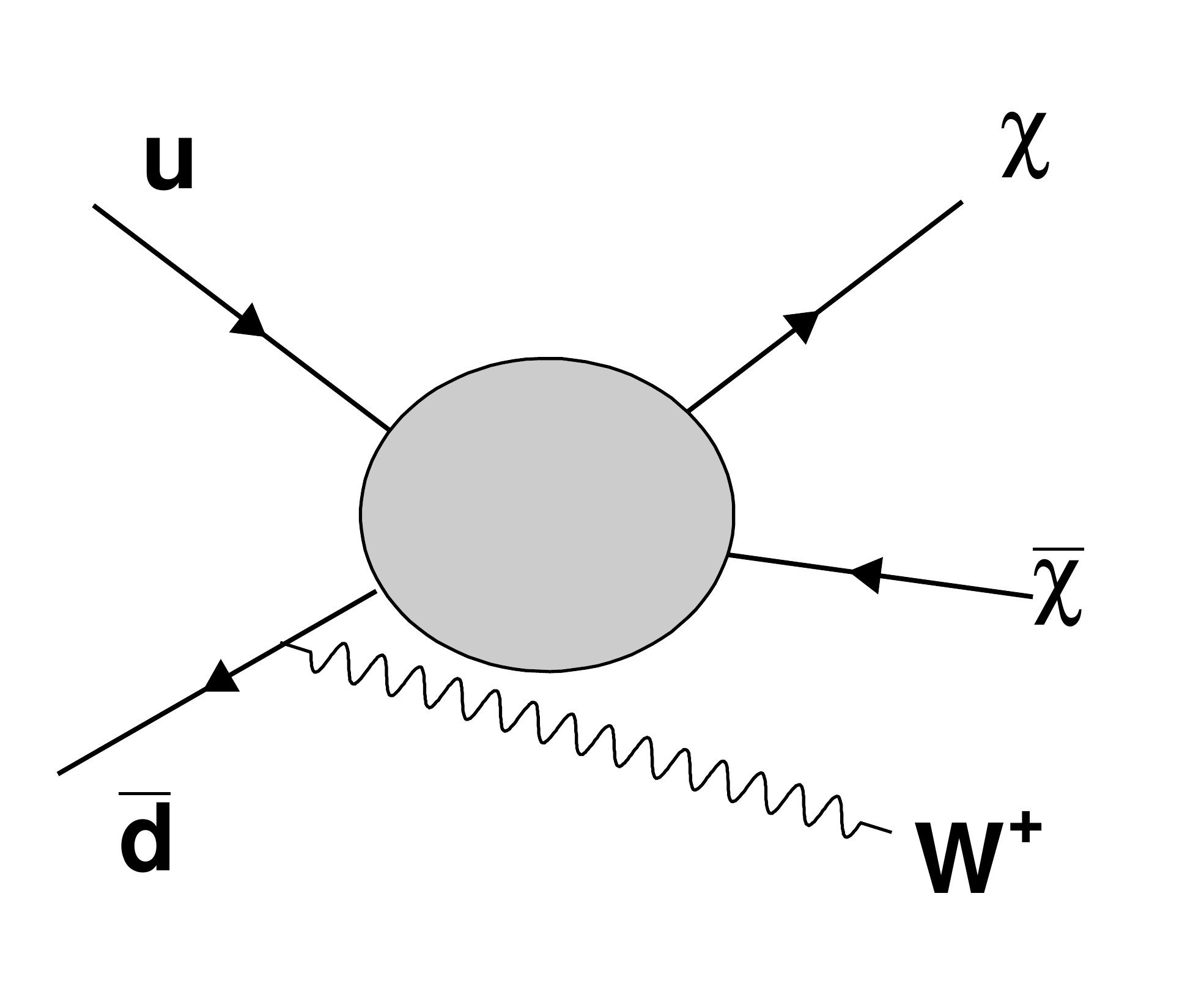}
\caption{ Pair production of WIMPs ($\chi\bar{\chi}$) in proton--proton
  collisions at the LHC via an unknown intermediate state, with
  initial-state radiation of a $W$ boson.}
\label{fig:monow}
\end{figure}

One critical component of this program is the search for
pair production of WIMPs at particle colliders, specifically
$pp\rightarrow \chi\bar{\chi}$ at the Large Hadron Collider (LHC) via some unknown
intermediate state. These searches  have greatest sensitivity at low WIMP mass $m_\chi$,
where direct detection experiments are less powerful. At the LHC, the final-state WIMPs are invisible to the
detectors, but the events can  be detected if there is associated
initial-state radiation of a SM particle~\cite{cdfjmet}; an
example is shown in Fig.~\ref{fig:monow}.

The Tevatron and LHC collaborations have reported limits on the cross section of
$pp\rightarrow  \chi\bar{\chi}+X$ where $X$ is a
hadronic jet~\cite{cdfjmet,atlasjet,cmsjet} or a
photon~\cite{atlasphoton,cmsphoton}. Other LHC data have been
reinterpreted to constrain models where $X$ is a leptonically decaying
$W$~\cite{monow} or $Z$ boson~\cite{atlaszz,monoz}. In each case,
limits are reported in terms of the mass scale
$M_*$ of the unknown interaction expressed in an effective field
theory as a four-point contact interaction~\cite{Cao:2009uw,Beltran:2010ww,Goodman:2010yf,Bai:2010hh,Goodman:2010ku,Rajaraman:2011wf,Fox:2011pm,Cheung:2012gi,Cotta:2012nj}.
In the models considered until now, the strongest limits come from monojet analyses, due to the large rate
of gluon or quark initial-state radiation relative to photon, $W$ or $Z$ boson
radiation.  The operators studied in 
these monojet and monophoton searches assume equal couplings of the dark matter
particles to up-type and down-type quarks [$C(u)=C(d)$]. For $W$ boson
radiation there is interference between the diagrams in which the $W$
boson is radiated from the $u$ quark or the $d$ quark. In the case of equal
coupling, the interference is destructive and gives a small $W$ boson emission
rate. If, however, the up-type and down-type couplings have opposite signs [$C(u)=-C(d)$] to give constructive interference, the
relative rates of gluon, photon, $W$ or $Z$ boson emission can change
dramatically~\cite{monow}, such that mono-$W$-boson production is the
dominant process.
  
In this Letter,  a search is reported for the production of $W$ or $Z$
bosons decaying hadronically (to $q\bar{q}'$ or $q\bar{q}$, respectively)
and reconstructed as a single massive jet in
association with large missing transverse momentum from the undetected
$\chi\bar{\chi}$ particles.  This search, the first
 of its kind, is sensitive to WIMP pair production,
as well as to other dark-matter-related models, such
as invisible Higgs boson decays ($WH$ or $ZH$ production with
$H\rightarrow\chi\bar{\chi}$).

The ATLAS detector~\cite{atlas} at the LHC covers the
pseudorapidity~\cite{coord} range $|\eta| <
4.9$ and the full azimuthal angle $\phi$. It consists of an inner tracking detector
surrounded by a thin superconducting solenoid, electromagnetic and
hadronic calorimeters, and an external muon spectrometer incorporating
large superconducting toroidal magnets. A three-level trigger system
is used to select interesting events for recording and subsequent
offline analysis. Only data for which beams were stable and all subsystems described above
were operational are used. Applying these requirements to $pp$ collision
data, taken at a center-of-mass energy of $\sqrt{s} = 8$~TeV during the 2012 LHC run, results in a data
sample with a time-integrated luminosity of 20.3~fb$^{-1}$. The systematic uncertainty on the luminosity is derived, following the same methodology as that detailed in Ref.~\cite{lumi}, from a preliminary calibration of the luminosity scale obtained from beam-separation scans performed in November 2012.

\begin{figure} [htbp]
  \center
  \includegraphics[width=0.85\linewidth]{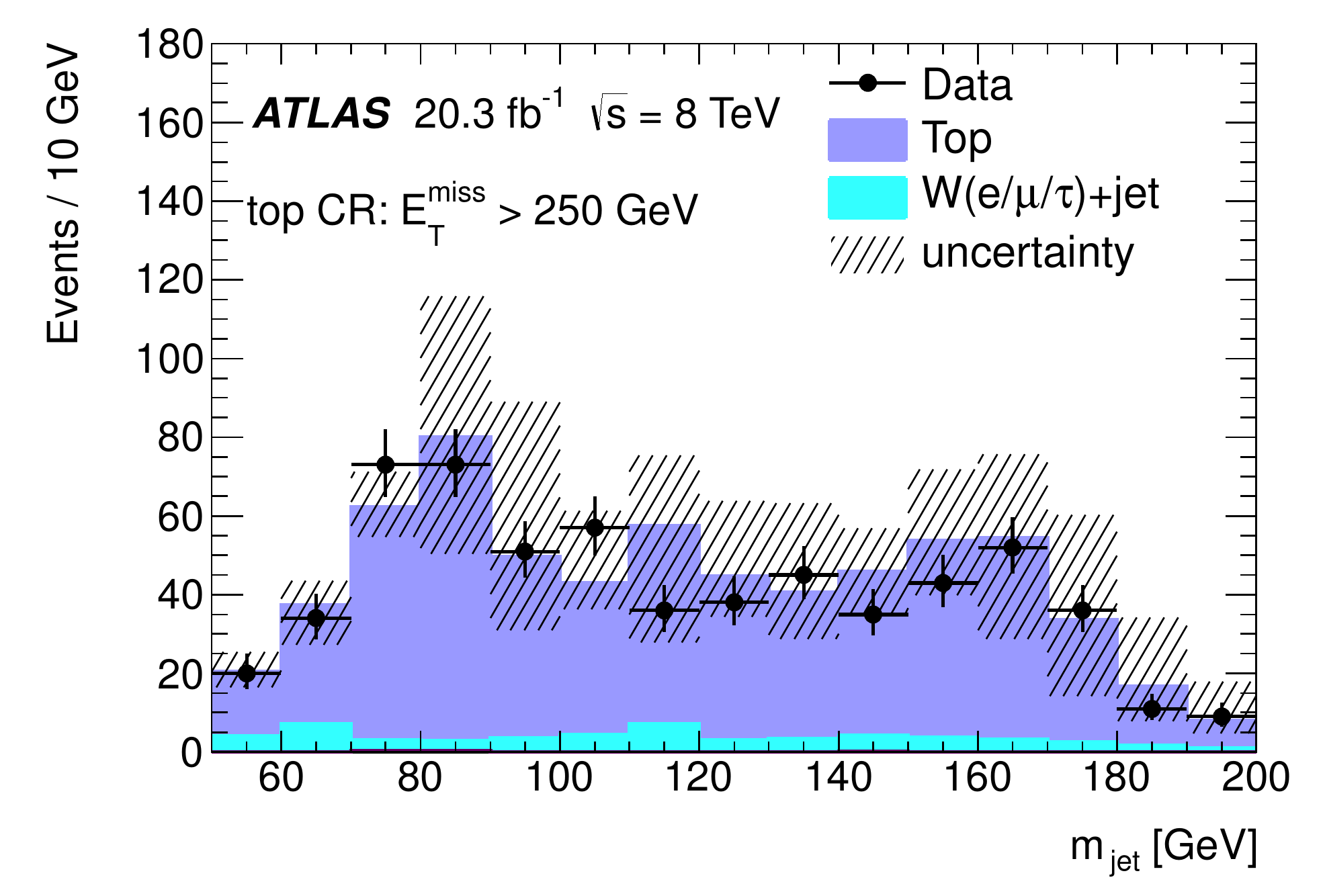}
  \caption{Distribution of $m_{\textrm{jet}}$ in the data and for the predicted background in the top control region (CR) with one muon, one large-radius jet, two narrow jets, at least one $b$ tag, and $\missET>250$~GeV, which includes a $W$ peak and a tail due to the inclusion of (part of) the $b$ jet from top decay.
    Uncertainties include statistical and systematic sources.}

  \label{fig:control}
\end{figure}

Jet candidates are reconstructed using the Cambridge--Aachen
algorithm~\cite{cajet} with a radius parameter of 1.2, and selected using a mass-drop filtering procedure~\cite{fatjetpub,bdrs}, referred to as {\it large-radius jets}. These large-radius jets are supposed to capture the
hadronic  products of both quarks from $W$ or $Z$ boson decay. The internal structure of the large-radius jet
is characterized in terms of the momentum balance of the two leading
subjets, as $\sqrt{y}= \textrm{min}(p_{\textrm T1},p_{\textrm
  T2})\Delta R/m_{\textrm{jet}}$ where $\Delta R =
\sqrt{(\Delta\phi_{1,2})^2+(\Delta\eta_{1,2})^2}$  and $m_{\textrm{jet}}$ is the calculated mass
of the jet. Jet
candidates are also reconstructed using the anti-$k_t$ clustering algorithm~\cite{Cacciari:2008gp}
with a radius parameter of 0.4, referred to as {\it narrow jets}. The inputs to both algorithms are
clusters of energy deposits in calorimeter cells seeded by those with
energies significantly above the measured noise and calibrated at the
hadronic energy scale~\cite{topocal}. Jet momenta are
calculated by performing a four-vector sum over these 
clusters, treating each topological cluster~\cite{topocal} as an $(E, \vec{p})$ four vector with zero
mass. The direction of $\vec{p}$ is given by the line joining the
reconstructed interaction point with the energy cluster. Missing
transverse momentum $\missET$ is measured using all clusters of
energy deposits in the calorimeter with $|\eta|<4.5$.  Electrons,
muons, jets, and $\missET$ are reconstructed as in
Refs~\cite{eledef,muodef,topocal,metdef}, respectively. The reconstruction of hadronic $W$ boson decays with
large-radius jets is validated in a $t\bar{t}$-dominated control region with one muon, one
large-radius jet ($p_{\textrm T}>250$~GeV, $|\eta|<1.2$), two additional narrow jets ($p_{\textrm T}>40$~GeV, $|\eta|<4.5$) separated from the leading large-radius jet, at least one $b$ tag, and $\missET>250$~GeV
(Fig.~\ref{fig:control}).

Candidate signal events are accepted by an inclusive $\missET$
trigger that is more than 99\% efficient for events with
$\missET>150$~GeV. Events with significant detector noise and
noncollision backgrounds are rejected as described in Ref.~\cite{atlasjet}.  In addition,
events are required to have at least one large-radius jet with $p_{\textrm T}>250$~GeV, $|\eta|<1.2$, $m_{\textrm{jet}}$ between 50~GeV
and 120~GeV, and $\sqrt{y}>0.4$ to suppress background without hadronic $W$ or $Z$ boson decays.  Two signal regions are defined by two
thresholds in $\missET$: 350 and 500~GeV.  To suppress the
$t\bar{t}$ background and multijet background, events are rejected if they
contain more than one narrow jet with $p_{\textrm T}>40$~GeV and
$|\eta|<4.5$ 
which is not completely overlapping with the leading large-radius jet by a separation of $\Delta R>0.9$,
or if any narrow jet has
$\Delta\phi(\missET,\textrm{jet})<0.4$.  Finally, to suppress
contributions from
$W\rightarrow \ell\nu$ production, events are rejected
if they have any electron, photon, or muon candidates with $p_{\textrm T}>10$~GeV and $|\eta|<$ 2.47, 2.37, or 2.5, respectively.

The dominant source of background events is $Z\rightarrow \nu\bar{\nu}$
production in association with jets from initial-state radiation. A secondary
contribution comes from production of jets in association with
$W$ or $Z$ bosons with leptonic decays 
in which the charged leptons fail identification requirements or the $\tau$ leptons decay hadronically. 
These three backgrounds are estimated by extrapolation from a common
data control region in which the selection is identical to that of the signal regions except that the muon veto is inverted and $W/Z+$jets with muon decays are the dominant processes.
In this muon control region dominated by $W/Z+$jets with muon decays, the combined $W$ and $Z$ boson contribution is measured after
subtracting other sources of background that are estimated using MC simulation~\cite{mcsim} based on {\sc geant\small4}~\cite{AGO-0301}.
Two extrapolation factors from the contribution of $W/Z+$jets in the
muon control region to the contributions of $Z\rightarrow\nu\nu$+jets and $W/Z+$jets with leptonic decays in the muon-veto signal region, respectively, are derived
as a function of $m_{\textrm{jet}}$ from simulated
samples of $W$ and $Z$ boson production in association with jets that are
generated using {\sc sherpa\small1.4.1}~\cite{sherpa} and the
CT10~\cite{ct10}  parton distribution function (PDF) set. 
A second control region is defined with two muons and $\missET>350$~GeV, which has limited statistics and is used only for the validation of the $Z$ boson contribution. The $W$ boson contribution is
validated in a low-$\missET$ control region with the same selection as the signal region but 250~GeV $<\missET<$ 350~GeV.

Other sources of background are diboson production, top quark pair
production, and single-top production, which are estimated using
simulated events.  The {\sc mc@nlo\small4.03} generator~\cite{mcnlo} using
 the CT10 PDF with the AUET2~\cite{auet1} tune, interfaced to {\sc herwig\small6.520}~\cite{herwig} and
{\sc jimmy\small4.31}~\cite{jimmy} for the simulation of underlying events,
is used for the productions of $t\bar{t}$ and  single-top
processes, both $s$-channel and $Wt$ production. The single-top, $t$-channel process is generated with
{\sc acermc\small3.8}~\cite{acermc} interfaced to
{\sc pythia\small8.1}~\cite{pythia8}, using the CTEQ6L1~\cite{cteq} PDF
with the AUET2B~\cite{auet1} tune. The diboson ($ZZ$, $WZ$, and $WW$) samples
are produced using {\sc herwig\small6.520} and {\sc jimmy\small4.31} with the
CTEQ6L1 PDF and AUET2 tune.

Background contributions from multijet production in which large
$\missET$ is due to mismeasured jet energies are estimated by
extrapolating from a sample of events with two jets and are found to be negligible~\cite{atlasjet}.

Samples of simulated $pp\rightarrow W\chi\bar{\chi}$ and
$pp\rightarrow Z\chi\bar{\chi}$ events are generated
using {\sc madgraph\small5}~\cite{madgraph}, with showering and hadronization modeled by
{\sc pythia\small8.1} using the AU2~\cite{auet1} tune and CT10 PDF, including $b$ quarks 
 in the initial state. 
Four operators are used as a representative set based on the definitions in Ref.~\cite{Goodman:2010ku}: $C1$ scalar, $D1$ scalar, $D5$ vector (both the constructive and destructive interference cases), and $D9$ tensor.
In each case,  $m_\chi=1,50,100,200,400,700,1000$, and 1300~GeV are used.
The dominant sources of systematic uncertainty are due to the limited
number of events in
  the control region, theoretical uncertainties in the
simulated samples used for extrapolation, uncertainties in the large-radius
jet energy calibration and momentum resolution~\cite{fatjetpub}, and uncertainties in the $\missET$. Additional minor uncertainties are due
to the levels of initial-state and final-state radiation, parton
distribution functions, lepton reconstruction and identification efficiencies, and momentum resolution.

\begin{figure} [htbp]
  \center
  \includegraphics[width=0.85\linewidth]{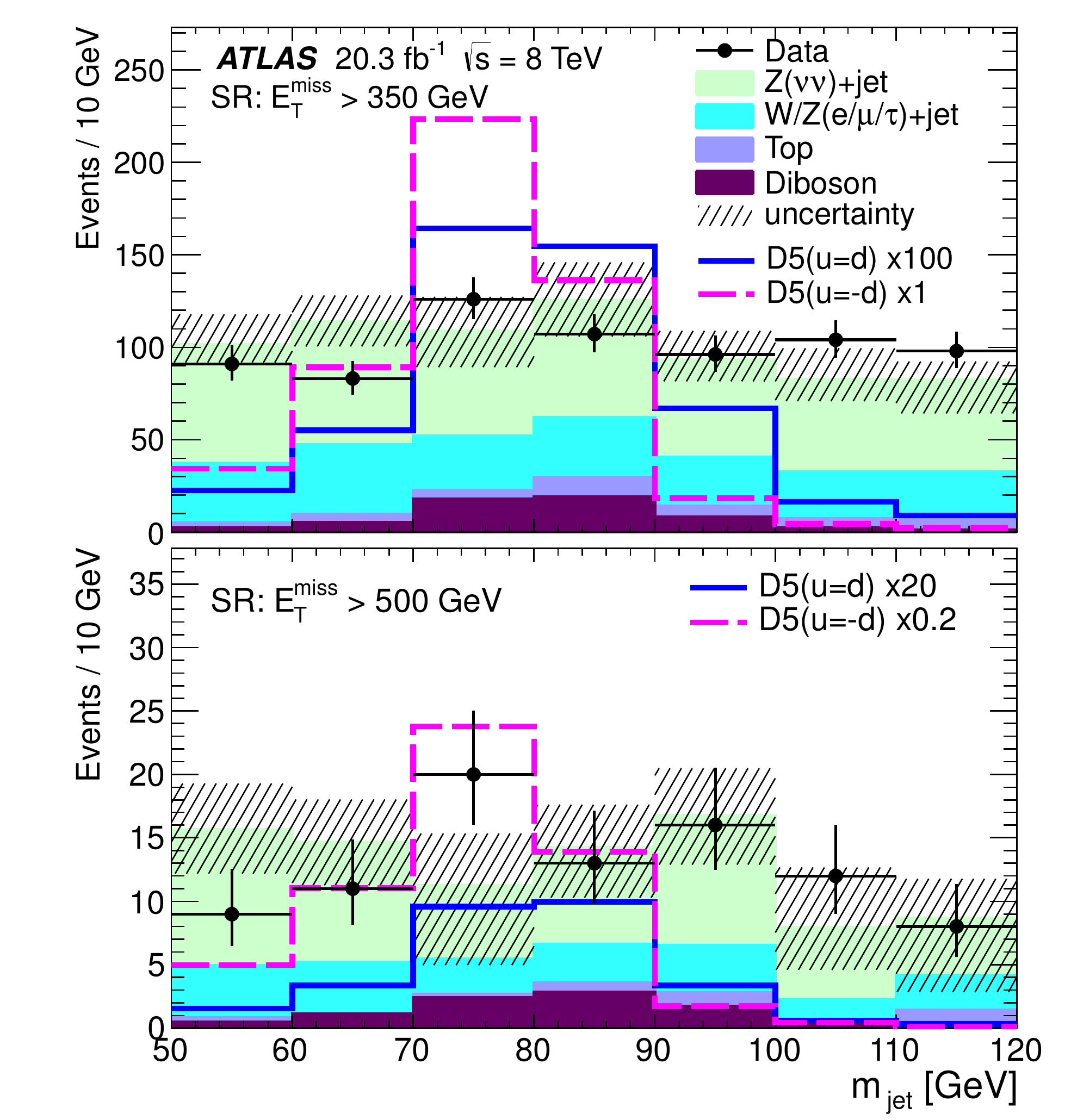}
  \caption{Distribution of $m_{\textrm{jet}}$ in the data and for the predicted background in the signal regions (SR) with $\missET>350$~GeV (top) and $\missET>500$~GeV (bottom). Also shown are the combined mono-$W$-boson and mono-$Z$-boson signal distributions with $m_{\chi}=1$~GeV and $M_*=1$~TeV for the $D5$ destructive and $D5$ constructive cases, scaled by factors defined in the legends.
Uncertainties include statistical and systematic contributions. 
}
  \label{fig:signalunblind}
\end{figure}

\begin{figure}[htbp]
  \center
  \includegraphics[width=0.85\linewidth]{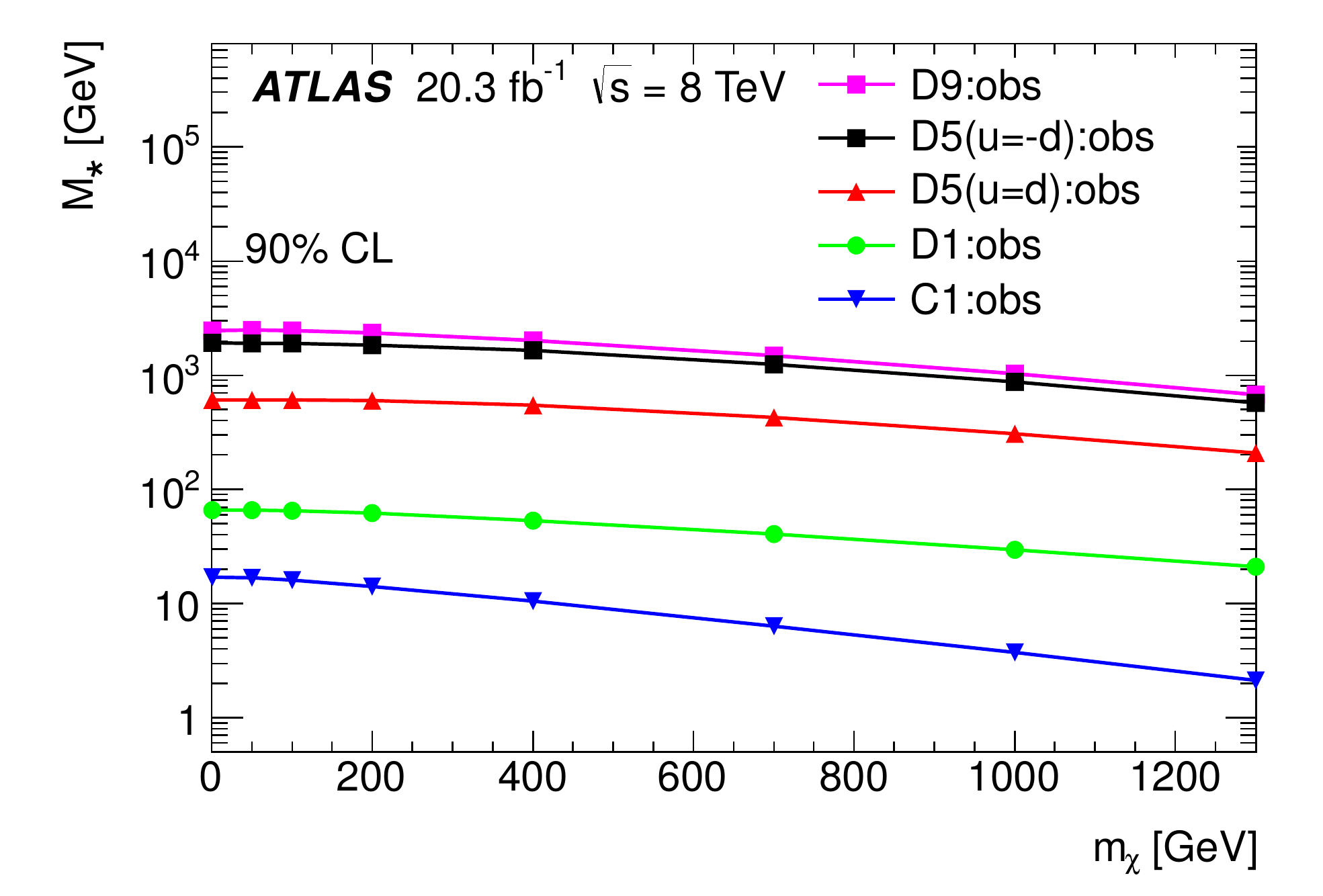}
  \caption{ Observed limits on the effective theory
    mass scale $M_*$ as a function of $m_{\chi}$ at 90\% C.L. 
    from combined mono-$W$-boson and mono-$Z$-boson signals for various operators. 
    For each operator, the values below the corresponding line are excluded. }
  \label{fig:limitsummarytop}
\end{figure}

The data and predicted backgrounds in the two signal regions are shown in
Table~\ref{tab:signalunblind} for the total number of events and in Fig.~\ref{fig:signalunblind} for the $m_{\textrm{jet}}$ distribution.  The data agree well with the background estimate for each $\missET$ threshold. Exclusion limits are set on
the dark matter signals using the predicted shape of the
$m_{\textrm{jet}}$ distribution and the $CL_{s}$ method~\cite{cls},
calculated with toy simulated experiments in which the systematic
uncertainties have been marginalized. Figure~\ref{fig:limitsummarytop} shows the exclusion regions at $90\%$ confidence level (C.L.) in the $M_*$ vs $m_\chi$ plane for various operators, where $M_*$ need not be the same for the different operators.

\begin{table}[tbh]
\caption{Data and estimated   background yields in the two signal
  regions. Uncertainties include statistical and systematic contributions.}
\center
\begin{tabular}{l|c|c}
\hline\hline
Process & $\missET>350$~GeV & $\missET>500$~GeV \\\hline
$Z\rightarrow\nu\bar{\nu}$                                    & $402 ^{+39}_{-34}$ & $54 ^{+8}_{-10}$ \\
$W\rightarrow \ell^\pm\nu$, $Z\rightarrow \ell^\pm\ell^\mp$   & $210 ^{+20}_{-18}$ & $22 ^{+4}_{-5}$ \\
$WW,WZ,ZZ$                                                    & $57 ^{+11}_{-8}$ & $9.1^{+1.3}_{-1.1}$ \\
$t\bar{t}$, single $t$                                        & $39 ^{+10}_{-4}$ & $3.7^{+1.7}_{-1.3}$ \\ \hline
Total                                                         & $707 ^{+48}_{-38}$ & $89 ^{+9}_{-12}$\\\hline
Data & 705 & 89 \\
\hline  \hline
\end{tabular}
\label{tab:signalunblind}
\end{table}

\begin{figure}[htbp]
  \center
  \includegraphics[width=0.95\linewidth]{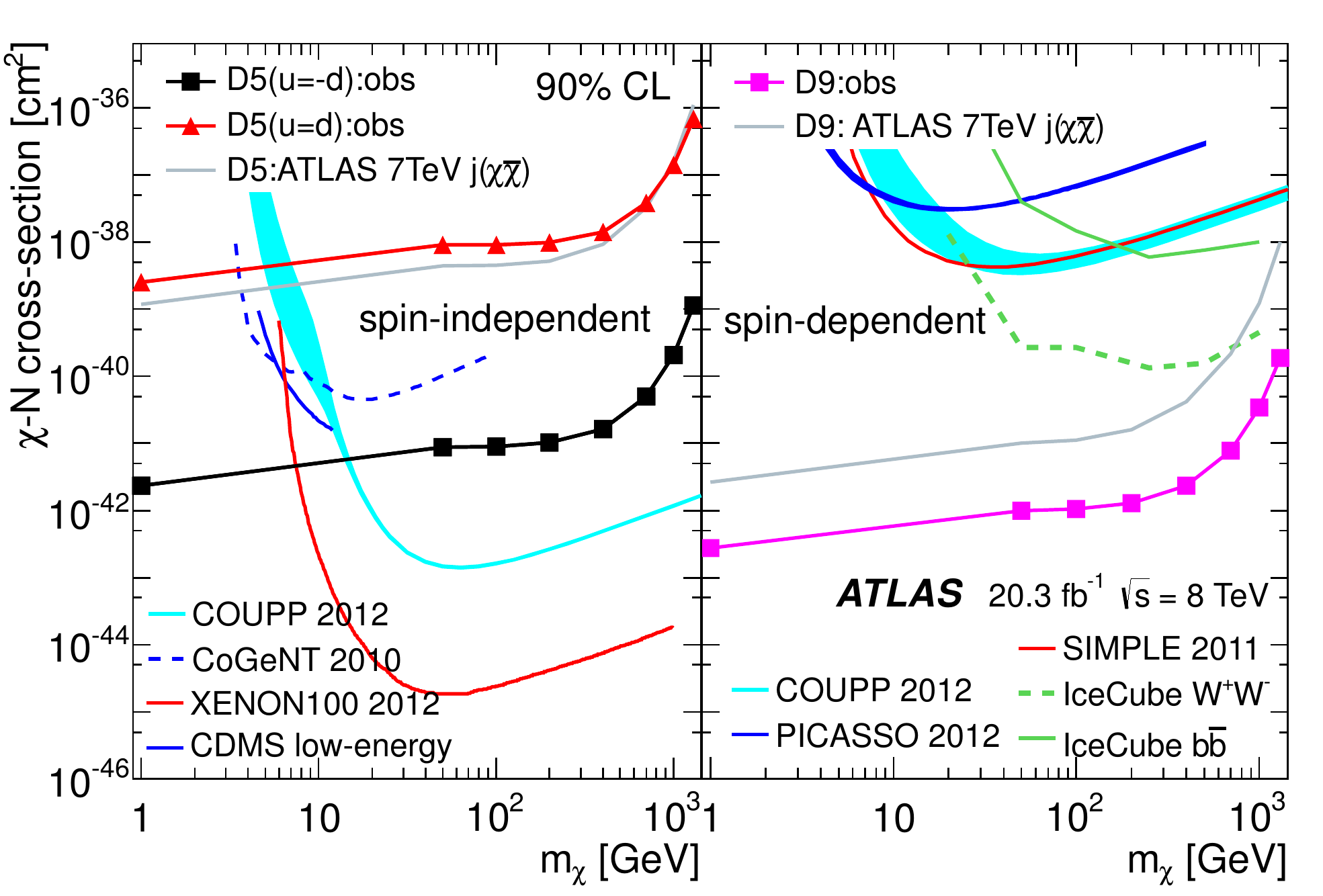}
  \caption{Limits on $\chi$--nucleon cross sections as a function of $m_{\chi}$ at $90\%$ C.L. for spin-independent (left) and spin-dependent (right) operators in effective field theory, compared to previous
    limits~\cite{xenon100,cdms,cogent,picasso,simple,icecube,coupp2012}.
  }
  \label{fig:limit_dd}
\end{figure}

\begin{figure}[htbp]
  \center
  \includegraphics[width=0.85\linewidth]{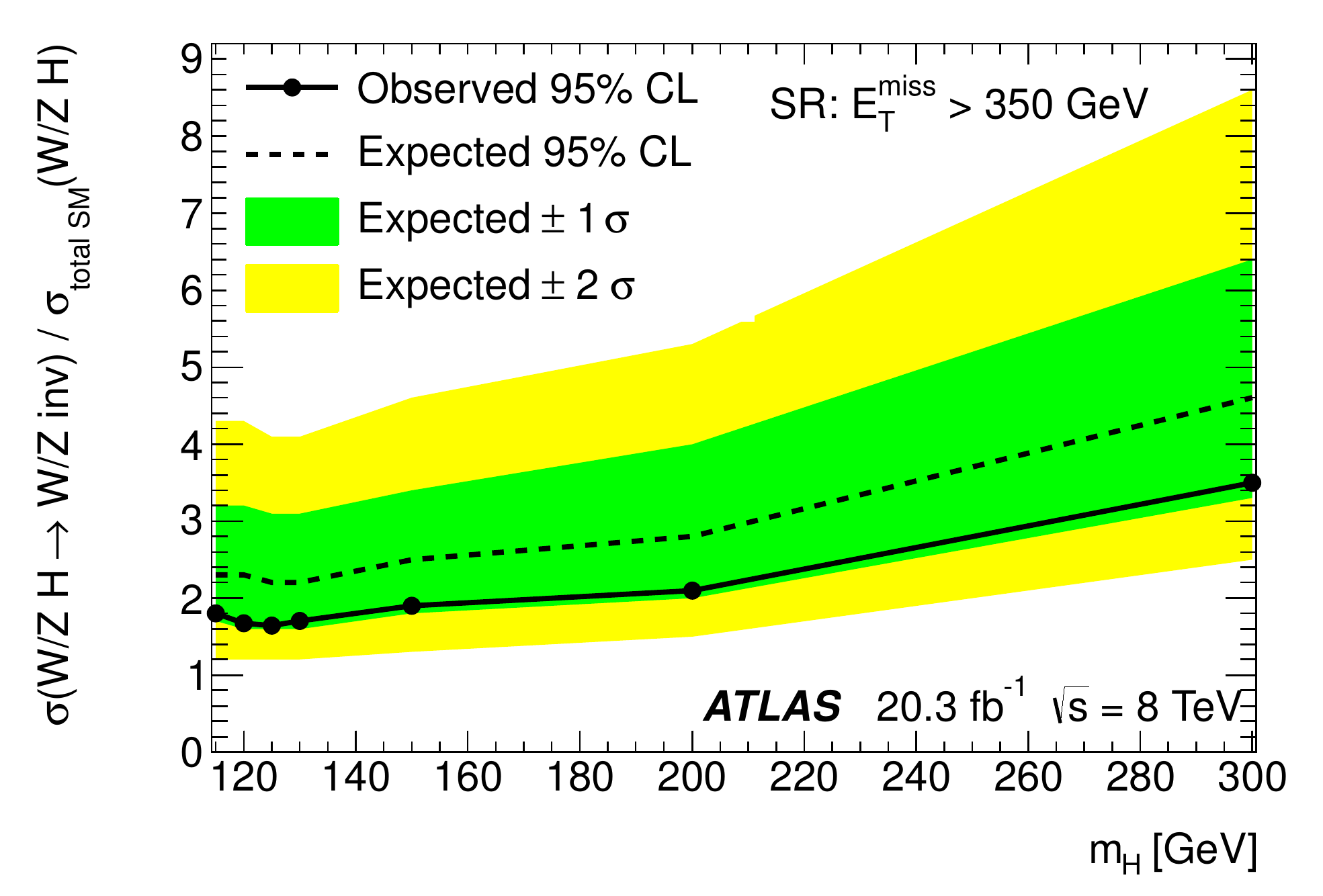}
  \caption{ Limit on the Higgs boson cross section for
    decay to invisible particles divided by the cross section for decays to
    standard model particles as a function of $m_{\textrm{H}}$ at 95\% C.L., derived from the signal region (SR) with $\missET>350$~GeV.}
  \label{fig:limitsummarybottom}
\end{figure}

Limits on the dark matter--nucleon scattering cross sections are
reported using the method of Ref.~\cite{Goodman:2010ku}  in Fig.~\ref{fig:limit_dd} for both the
spin-independent ($C1$, $D1$, $D5$) and the spin-dependent
interaction model ($D9$).  References~\cite{Goodman:2010ku,Busoni:2013lha} discuss the valid region of the effective field theory, which becomes a poor approximation if the mass of the intermediate state is below the momentum transferred in the interaction. 
The results are compared with measurements from direct detection experiments~\cite{xenon100,cdms,cogent,picasso,simple,icecube,coupp2012}.  

This search for dark
matter pair production in association with a $W$ or $Z$ boson extends
 the limits on the dark matter--nucleon scattering cross
section in the low mass region $m_{\chi}<10$~GeV where the direct detection experiments have less sensitivity. The new limits are also compared to the limits set by ATLAS in the $7$~TeV monojet analysis~\cite{atlasjet}. For the spin-independent case with the opposite-sign up-type and down-type couplings, the limits are improved by about 3 orders of magnitude,
as the constructive interference leads to a very large increase in the $W$-boson-associated production cross section.
For other cases, the limits are similar.

To complement the effective field theory models, limits are calculated
for a simple dark matter production theory with a light mediator, the Higgs boson.  The
upper limit on the cross section of Higgs boson production through
$WH$ and $ZH$ modes and decay to invisible particles is 1.3~pb at 95\%
C.L. for $m_H=125$ GeV. Figure~\ref{fig:limitsummarybottom} shows the upper limit
of the total cross section of $WH$ and $ZH$ processes with $H\rightarrow\chi\bar{\chi}$, normalized to the SM next-to-leading order prediction for the $WH$ and $ZH$ production cross section (0.8~pb for $m_H=125$ GeV)~\cite{higgsnlo}, which is 1.6 at $95\%$ C.L. for $m_H=125$ GeV.

In addition, limits are calculated on dark matter $W\chi\bar{\chi}$ or $Z\chi\bar{\chi}$ production within two
fiducial regions defined at parton level: $p^{W\textrm{or}
  Z}_{\textrm{T}} > 250$~GeV, $|\eta^{W\textrm{or} Z}| < 1.2$; two
quarks from $W$ or $Z$ boson decay with $\sqrt{y}>0.4$; at most one additional narrow jet [$p_{\textrm{T}}>40$~GeV, $|\eta|<4.5$, $\Delta R(\textrm{narrow~jet}, W~\textrm{or}~Z)>0.9$]; no electron, photon, or muon with $p_\textrm{T}>10$~GeV and $|\eta|<$ 2.47, 2.37, or 2.5, respectively; $p_{\textrm{T}}^{\chi\bar{\chi}}>350$ or $500$~GeV. 
The fiducial efficiencies are similar for various dark matter signals, and the smallest value is $(63\pm1)\%$ in both fiducial regions.
The observed upper limit on the fiducial cross section is \mbox{4.4 fb} \mbox{(2.2 fb)} at $95\%$ C.L. 
for $p_{\textrm{T}}^{\chi\bar{\chi}}>350$~GeV ($500$~GeV) and the expected limit is \mbox{5.1 fb} \mbox{(1.6 fb)} with negligible dependence on the dark matter production model.

In conclusion, this Letter reports the first LHC limits on dark matter
production in events with a hadronically decaying $W$  or
$Z$ boson and large missing transverse  momentum. In the case of
constructive interference between up-type and down-type
contributions, the results set the strongest limits on the mass scale
of $M_*$ of the unknown mediating interaction, surpassing those from
the monojet signature.

\input{Acknowledgements}

\bibliography{monow}
\onecolumngrid
\clearpage
\onecolumngrid
\input{atlas_authlist}

\end{document}

%% file: Acknowledgements.tex
%\documentclass[11pt,a4paper,dvips]{article}

%\begin{document}

% Acknowledgements for papers with collision data
% Version 14-Nov-2013

%\section{Acknowledgements}

% Standard acknowledgements start here
%----------------------------------------------
We thank CERN for the very successful operation of the LHC, as well as the
support staff from our institutions without whom ATLAS could not be
operated efficiently.

We acknowledge the support of ANPCyT, Argentina; YerPhI, Armenia; ARC,
Australia; BMWF and FWF, Austria; ANAS, Azerbaijan; SSTC, Belarus; CNPq and FAPESP,
Brazil; NSERC, NRC, and CFI, Canada; CERN; CONICYT, Chile; CAS, MOST, and NSFC,
China; COLCIENCIAS, Colombia; MSMT CR, MPO CR, and VSC CR, Czech Republic;
DNRF, DNSRC, and Lundbeck Foundation, Denmark; EPLANET, ERC, and NSRF, European Union;
IN2P3-CNRS, CEA-DSM/IRFU, France; GNSF, Georgia; BMBF, DFG, HGF, MPG, and AvH
Foundation, Germany; GSRT and NSRF, Greece; ISF, MINERVA, GIF, DIP, and Benoziyo Center,
Israel; INFN, Italy; MEXT and JSPS, Japan; CNRST, Morocco; FOM and NWO,
Netherlands; BRF and RCN, Norway; MNiSW and NCN, Poland; GRICES and FCT, Portugal; MNE/IFA, Romania; MES of Russia and ROSATOM, Russian Federation; JINR; MSTD,
Serbia; MSSR, Slovakia; ARRS and MIZ\v{S}, Slovenia; DST/NRF, South Africa;
MINECO, Spain; SRC and Wallenberg Foundation, Sweden; SER, SNSF, and Cantons of
Bern and Geneva, Switzerland; NSC, Taiwan; TAEK, Turkey; STFC, the Royal
Society, and Leverhulme Trust, United Kingdom; U.S. DOE and NSF, United States of
America.

The crucial computing support from all WLCG partners is acknowledged
gratefully, in particular from CERN and the ATLAS Tier-1 facilities at
TRIUMF (Canada), NDGF (Denmark, Norway, Sweden), CC-IN2P3 (France),
KIT/GridKA (Germany), INFN-CNAF (Italy), NL-T1 (Netherlands), PIC (Spain),
ASGC (Taiwan), RAL (UK) and BNL (U.S.), and in the Tier-2 facilities
worldwide.
%----------------------------------------------
%\end{document}

%% file: atlas_authlist.tex
% ATLAS Collaboration author list
% Data extracted on 12-Nov-2013 for paper reference EXOT-2012-27
%\documentclass[11pt]{article}
%\usepackage{a4wide}\begin{document}
\begin{flushleft}
{\Large The ATLAS Collaboration}

\bigskip

G.~Aad$^{\rm 48}$,
T.~Abajyan$^{\rm 21}$,
B.~Abbott$^{\rm 112}$,
J.~Abdallah$^{\rm 12}$,
S.~Abdel~Khalek$^{\rm 116}$,
O.~Abdinov$^{\rm 11}$,
R.~Aben$^{\rm 106}$,
B.~Abi$^{\rm 113}$,
M.~Abolins$^{\rm 89}$,
O.S.~AbouZeid$^{\rm 159}$
% ...
\bigskip
\\
% ...
$^{11}$ Institute of Physics, Azerbaijan Academy of Sciences, Baku, Azerbaijan\\
$^{12}$ Institut de F{\'\i}sica d'Altes Energies and Departament de F{\'\i}sica de la Universitat Aut{\`o}noma de Barcelona, Barcelona, Spain\\
% ...
$^{21}$ Physikalisches Institut, University of Bonn, Bonn, Germany\\
% ...
$^{48}$ Fakult{\"a}t f{\"u}r Mathematik und Physik, Albert-Ludwigs-Universit{\"a}t, Freiburg, Germany\\
% ...
$^{89}$ Department of Physics and Astronomy, Michigan State University, East Lansing MI, United States of America\\
% ...
$^{106}$ Nikhef National Institute for Subatomic Physics and University of Amsterdam, Amsterdam, Netherlands\\
% ...
$^{112}$ Homer L. Dodge Department of Physics and Astronomy, University of Oklahoma, Norman OK, United States of America\\
$^{113}$ Department of Physics, Oklahoma State University, Stillwater OK, United States of America\\
% ...
$^{116}$ LAL, Universit{\'e} Paris-Sud and CNRS/IN2P3, Orsay, France\\
% ...
$^{159}$ Department of Physics, University of Toronto, Toronto ON, Canada\\
% ...
\end{flushleft}

%\end{document}
% Created with ./xml2latex.py

%% file: monow.bbl
\begin{thebibliography}{51}
\bibitem[{coo()}]{coord}
\bibinfo{howpublished}{ATLAS uses a right-handed coordinate system with its
  origin at the nominal interaction point (IP) in the center of the detector
  and the $z$-axis along the beam pipe. The $x$-axis points from the IP to the
  center of the LHC ring, and the $y$-axis points upward. Polar coordinates
  ($r, \phi$) are used in the transverse ($x,y$)-plane, $\phi$ being the
  azimuthal angle around the beam pipe. The pseudorapidity is defined in terms
  of the polar angle $\theta$ as $\eta = -\ln \tan(\theta/2)$.}
\end{thebibliography}
